\documentclass[prd,nofootinbib,showpacs]{revtex4}
\usepackage{amsmath,amssymb,graphicx,epsfig,color}

\begin{document}

\title{$B$-to-Glueball form factor and Glueball production in $B$ decays}
\author{ Wei Wang$^{a}$, Yue-Long Shen$^{b,c}$ and Cai-Dian L\"u$^{a}$}
\affiliation{
 $^a$ Institute of High Energy Physics and Theoretical Physics Center for Science Facilities,
  Chinese Academy
of Sciences, Beijing 100049, People's Republic of China \\
\it $^b$ Institute of Physics, Academia Sinica, Taipei, Taiwan115, Republic of China\\
 \it $^c$ College of Information Science and Engineering,
Ocean University of China, Qingdao, Shandong 266100, People's Republic of China
}

\begin{abstract}
We investigate transition form factors of $B$ meson decays into a
scalar glueball in the light-cone formalism. Compared with form
factors of $B$ to ordinary scalar mesons, the $B$-to-glueball form
factors have the same power in the expansion of $1/m_B$. Taking into
account the leading twist light-cone distribution amplitude, we find
that they are numerically  smaller than those form factors of $B$ to
ordinary scalar mesons. Semileptonic $B\to Gl\bar\nu$, $B\to
Gl^+l^-$ and $B_s\to Gl^+l^-$ decays are subsequently investigated.
We also analyze the production rates of scalar mesons in
semileptonic $B$ decays in the presence of mixing between scalar
$\bar qq$ and glueball states. The glueball production in $B_c$
meson decays is also investigated and the LHCb experiment may
discover this channel. The sizable branching fraction in $B_c\to
(\pi^+\pi^-)l^-\bar\nu$, $B_c\to (K^+K^-)l^-\bar\nu$ or $B_c\to
(\pi^+\pi^-\pi^+\pi^-)l^-\bar\nu$ could be a clear signal for a
scalar glueball state.
\end{abstract}

 \maketitle

\section{Introduction}

The existence of the glueball state is permitted by the QCD. Based
on the valence approximation,  Lattice QCD calculations have
predicted that the mass of the lowest-lying scalar glueball is
around 1.5-1.8 GeV~\cite{Bali:1993fb,Chen:2005mg}. Many different
candidates have been observed in this mass region, but there is not
any solid evidence on the existence of a pure glueball. It is very
likely that the glueball mixes with the  quark-antiquark state and
they together form several physical mesons. On the theoretical side,
there are large ambiguities on the mixing mechanism, please see
Ref.~\cite{scalarmesons} for a review on the status.

Most studies focus on the decay property of the glueball. In fact,
the production property of the glueball is an alternative way  to
uncover the mysterious structure of scalar mesons and figure out the
gluon component inside. In $B$ meson decays, the color magnetic
operator $O_{8g}$ has a large Wilson coefficient that could produce
a number of gluons easily. These gluons in the final state may have
the tendency to form a glue state, thus the glueball production in
inclusive $B$ decays has attracted some theoretical
interests~\cite{Minkowski:2004xf,He:2006qk}. The authors in
Ref.~\cite{Chen:2007uq} have also studied the exclusive $B\to KG$
and $B\to K^*G$ decays, where $G$ is a scalar glueball. Based on the
results in Ref.~\cite{Charng:2006zj}, they have neglected gluon
recoiled Feynman diagrams and studied the contributions, in which a
scalar glueball state is emitted and a $K$ or $K^*$ meson is
recoiled.

In the present paper, we study the transition form factors of $B$
decays into a scalar glueball and point out another interesting
mechanism to detect a glueball via the exclusive $B$ decays. These
form factors are relevant for productions of scalar glueballs in
semileptonic $B$ decays. Up to the leading Fock state and leading
order in $\alpha_s$, there are three different Feynman diagrams
shown in Fig.~\ref{diagram:FeynBtof0}. Since the two gluons bounded
in a glueball state are already symmetrized in the glueball wave
function, it is not necessary to consider the crossed diagram. The
first diagram is similar with the one studied in
Ref.~\cite{Charng:2006zj}. Through the power counting rule, we will
show that the first two diagrams have the same power with the
$B$-to-light meson form factors. That will affect numerical results
of the transition form factors and production rates in semileptonic
$B$ decays.

The ordinary light neutral scalar meson is isospin singlet and/or
flavor SU(3) singlet, while the glueball is flavor SU(6) singlet.
Therefore it is difficult to distinguish them by the light $u$,
$d$ and $s$ quark coupling. However, the light ordinary scalar
meson has negligible $c\bar c$ component, while the glueball have
the same coupling to $c\bar c$ as that to the  $u \bar u$, $d \bar
d$ or $s \bar s$. A clean way to identify a glueball is then
through the $c\bar c$ coupling to the glueball. We briefly analyze
the production in $B_c$ meson decays.

The paper is organized as follows. In
section~\ref{sec:powercounting}, we give the analysis of the power
counting of the $B\to G$ transition form factors. In the
perturbative QCD (PQCD) approach~\cite{Keum:2000ph,Lu:2000em}, the
form factors are calculated in the
section~\ref{sec:PQCDtransitionformfactors}. Numerical results and
detailed  discussions will be presented in the subsequent section.
We also briefly analyze the discussion in $B_c$ meson decays.  We
conclude in the last section. Functions used in the PQCD approach
are collected in the appendix.

\section{Power counting of transition form factors}
\label{sec:powercounting}

Transition form factors of $B$ decays into scalar mesons are defined
by
  \begin{eqnarray}
   \langle S(P_S) |\bar q  \gamma_\mu\gamma_5  b|\overline  {B}(P_{B})\rangle
    &=&-i\left\{\left
    [(P_{B} +P_S)_\mu - \frac{m_{B}^2-m_S^2}{q^2}q_\mu \right ]
    F_1 (q^2)
    +\frac{m_{B}^2-m_S^2}{q^2}q_\mu F_0 (q^2)\right\} ,\nonumber\\
   \langle S(P_S) |\bar q  \sigma_{\mu\nu} b|\overline  {B}(P_{B})\rangle
   &=&i\epsilon_{\mu\nu\alpha\beta}P_S^{\alpha}q^{\beta}
  \frac{2 F_T (q^2)}{
   m_{B}+m_S},\nonumber\\
   \langle S(P_S) |\bar q  \sigma_{\mu\nu}\gamma_5  b|\overline  {B}(P_{B})\rangle
   &=& -
  \frac{F_T (q^2)}{
   m_{B}+m_S}[  (P_B+P_S)_\mu q_\nu -q_\mu (P_B+P_S)_\nu],
   \end{eqnarray}
where $S$ denotes an ordinary scalar meson or a scalar glueball.
$q=P_{B}-P_S$ and $m_B(m_S)$ is the mass of the $B$ (scalar) meson.

\begin{figure}
\begin{center}
\includegraphics[scale=0.5]{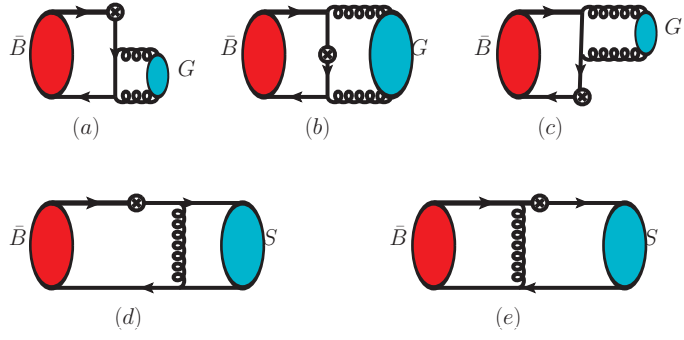}
\end{center}
\caption{Feynman diagrams of $\bar B$ decays into a scalar glueball
$G$ (first row) and an ordinary scalar meson (second row). The
$\otimes$ denotes the possible Lorentz structure arising from the
electroweak interactions.
} \label{diagram:FeynBtof0}
\end{figure}
%
%

In the following, we will work in the rest frame of the $B$ meson
and use light-cone coordinates. In the heavy quark limit, the mass
difference between $b$ quark ($m_b$) and $B$ meson ($m_B$) is
negligible. The mass of a light glueball is small compared with
the $b$ quark mass, thus in the transition amplitudes we keep them
up to the leading order. Since the scalar glueball in the final
state moves very fast in the large-recoil region, we choose its
momentum mainly on the plus direction in the light-cone
coordinates. The momentum of $B$ meson and the light scalar meson
can be denoted as
 \begin{eqnarray}
 P_{B}=\frac{m_{B}}{\sqrt{2}}(1,1,0_{\perp})\;,\;
 P_2=\frac{m_{B}}{\sqrt{2}}(\rho,0,0_{\perp})\;.\label{eq:momentum}
 \end{eqnarray}
The momentum transfer is $q=P_{B}-P_2$, and there exists the
relation $\rho\approx1-q^2/m_{B}^2$.

In the expansion on $\alpha_s$, the lowest order Feynman diagrams
for form factors of $B$ decays into a scalar glueball are depicted
in Fig.~\ref{diagram:FeynBtof0}. Up to the leading Fock state, two
gluons should be generated to form a glueball. In exclusive $B$
decays, these two gluons can be emitted from either the heavy $b$
quark or the light quark. The $\otimes$ in
Fig.~\ref{diagram:FeynBtof0} denotes the weak vertices. Momentum
fractions of the light antiquark in $B$ meson and the upper
(lower) gluon depicted in Fig~\ref{diagram:FeynBtof0} are denoted
as $x_1$ and $x_2$ ($\bar x_2=1-x_2$), respectively.

In the first diagram of Fig.~\ref{diagram:FeynBtof0}, each of the
two internal quark propagators is the sum of a collinear momentum
(gluon) and a soft momentum (light quark). The virtualities of
them are of order $\Lambda_{\rm QCD}m_b$, where $\Lambda_{\rm
QCD}$ is the hadronic scale. In the second and the third diagram,
one or two light propagators become heavy $b$ quark propagators,
whose virtualities
become $m_b^2$ instead of $\Lambda_{\rm QCD}m_b$. 
Superficially, one may conclude that the power counting for the
three diagrams obey the relation: $F(a)>F(b)>F(c)$. But in fact,
this relation is not exactly correct. As we will show in the
following, the leading twist light-cone distribution amplitude  is
constructed in the case that the two gluons are transversely
polarized. So the structures of the vertices attaching to these
two gluons in the first diagram have the form:
$\gamma_{\perp}^{\mu}$. Apparently, the numerator of the
propagators commutes with the transverse Dirac matrix. So, the
amplitude is proportional to $(\bar
x_2P\!\!\!\slash_2-x_1P\!\!\!\slash_B)
(P\!\!\!\slash_2-x_1P\!\!\!\slash_B)
\approx-x_1P\!\!\!\slash_BP\!\!\!\slash_2-x_1\bar
x_2P\!\!\!\slash_2P\!\!\!\slash_B$. We can see that neglecting the
glueball's mass square,  there is a small momentum fraction $x_1$
in the numerator, which will cancel one of the momentum fraction
of the denominator. The effective power for one light propagator
in the first diagram becomes $m_b^2$, which is the same as a heavy
propagator in the second diagram. It implies: $F(a)\sim
F(b)>F(c)$. Adopting the power counting rule for the $B$ meson
wave function and the scaling behaviors for the distribution
amplitudes of collinear meson,  given in
Ref.~\cite{Beneke:2000ry}, we directly obtain
\begin{eqnarray}
 F^{B\to G}\sim \alpha_s(\sqrt{m_b\Lambda_{\rm QCD}})\left(\frac{\Lambda_{\rm
 QCD}}{m_b}\right)^{3/2},
\end{eqnarray}
where the form factor is dominated by the first two diagrams. The
$B\to G$ form factors have the same scaling rule with the
$B$-to-light transition form factor. And we can expect that the
gluonic $B\to \eta$ form factors also obey this rule. While the
light-cone distribution amplitude of a gluonic pseudo-scalar meson
is normalized to zero, only the higher Gegenbauer moments
contribute. The first effective Gegenbauer moment of $\eta$ and
$\eta'$ is very small, so the gluonic contribution to $B\to \eta$
form factors is found to be numerically
small~\cite{Charng:2006zj}.

Since these diagrams are free of the endpoint singularity, both
collinear factorization  and $k_T$ factorization are applicable. In
the following, we will use the perturbative QCD approach to
investigate the transition form factors, where   the inclusion of
the transverse momentum only improves the theoretical precision.


\section{Transition form factors in the perturbative QCD approach}
\label{sec:PQCDtransitionformfactors}

$B$ meson is  a heavy-light system, whose light-cone matrix element
can be decomposed as~\cite{grozin,qiao}
 \begin{eqnarray}
 &&\int_0^1\frac{d^4z}{(2\pi)^4}e^{ik_1\cdot z}\langle 0|b_{\beta}(0)\bar
 q_{\alpha}(z)|\bar B(P_{B})\rangle=
 \frac{i}{\sqrt{2N_c}}\left\{(\not\! P_{B}+m_{B})\gamma_5\left[\phi_{B}(k_1)+\frac{\not\!
 n }{\sqrt{2}}\bar\phi_{B}(k_1)\right]\right\}_{\beta\alpha}\;,\label{Bwav:decompose}
 \end{eqnarray}
where $n=(1,0,\textbf{0}_T)$ and $v=(0,1,\textbf{0}_T)$ are
light-like unit vectors. There are two Lorentz structures in $B$
meson light-cone distribution amplitudes, and they obey the
normalization conditions:
 \begin{equation}
 \int\frac{d^4 k_1}{(2\pi)^4}\phi_{B}({
 k_1})=\frac{f_{B}}{2\sqrt{2N_c}}\;,\; \int \frac{d^4
 k_1}{(2\pi)^4}\bar{\phi}_{B}({ k_1})=0,\label{Bwave:normalization}
 \end{equation}
with $f_{B}$ as the decay constant of $B$ meson.

In principle, both the $\phi_{B}(k_1)$ and $\bar\phi_{B}(k_1)$
contribute in $B$ meson decays.  For $B$-to-glueball transition, the
$\bar \phi_{B}(k_1)$ can be neglected in the second diagram. The
amplitude for the form factors is obtained by evaluating the trace
of the wave functions and the scattering kernels. As we will show in
the following, the gluon in a glueball is transversely polarized at
the leading power. So the Lorentz structure $n\!\!\!\slash$ for the
wave function $\bar \phi_{B}(k_1)$ commutes with the gluon vertex
and encounters the propagator $(\bar
x_2P\!\!\!\slash_2-x_1P\!\!\!\slash_B)$. The momentum of the
glueball is parallel with this direction $n$, thus the nonzero term
contains a factor $x_1$ and is naturally power suppressed. For the
other diagrams, we do not find any analytical reason to neglect the
distribution amplitude $\bar\phi_{B}(k_1)$. The computation depends
on the shape of this distribution amplitude but it is not well
constrained in the PQCD approach at present. In fact, including the
$\bar\phi_{B}(k_1)$ term will not improve the quality of the
calculation significantly, but introduce one more free parameter.
Nevertheless in $B$ decays into ordinary $\bar qq$ mesons the
contribution of $\bar \phi_{B}(k_1)$ is usually neglected, because
its contribution is numerically small in the PQCD
approach~\cite{Wei:2002iu,Lu:2002ny}. We will keep the term with
$\phi_{B}(k_1)$ in equation (\ref{Bwav:decompose}) in this work and
leave the term with $\bar\phi_{B}(k_1)$ in a future study. In the
momentum space the light cone matrix of $B$ meson can be expressed
as:
 \begin{eqnarray}
 \Phi_{B}=\frac{i}{\sqrt{6}}(\not \! P_{B} +m_{B})\gamma_5\phi_{B} (k_1). \label{Bwave:3variable}
 \end{eqnarray}
Usually the hard part is independent of $k^+$ or/and $k^-$, so we
integrate one of them out from
$\phi_{B}(k^+,k^-,\textbf{k}_{\perp})$.

For the light-cone wave functions of the $B$ meson, we   use
 \begin{equation}
 \Phi_{{B},\alpha\beta}(x,b) = \frac{i}{\sqrt{2N_c}}
 \left[\not\! P_{B} \gamma_5 + m_{B}\gamma_{5} \right]_{\alpha\beta}
 \phi_{B}(x,b),\label{Bwave:bspace}
 \end{equation}
where $x$ is the momentum fraction of the light quark in $B$ meson
and $b=|{\bf b}|$ , which is the conjugate space coordinate of $
{\bf {k}_{\perp}}$. In this paper, we use the following model for
$\phi_{B}(x,b)$:
 \begin{equation}
 \phi_{B}(x,b)=N_{B}x^2(1-x)^2\mbox{exp}\left[-\frac{m_{B}^2 x^2}{2\omega_b^2}-\frac{1}{2}(\omega_b
 b)^2\right]. \label{Bwave:da}
 \end{equation}
The normalization factor $N_{B}$ is determined by normalization
condition. 

Decay constant of a scalar glueball state is defined as:
\begin{eqnarray}
 \langle G(P)|F^{\mu\rho}F^{\nu}_{\;\;\rho}|0\rangle &=& f_1
 m_{f_0}^2g^{\mu\nu}+f_s P^\mu P^\nu,\label{eq:decayconstant}
\end{eqnarray}
where $F_{\mu\nu}$ is the gluon field strength tensor.  $f_s$ is
determined as $f_s= (100\sim 130)$ MeV~\cite{He:2002hr}. When the
two gluons are separated in coordinate space, the nonperturbative
matrix elements are parameterized in terms of the light cone
distribution amplitudes (LCDAs). According to the conformal
symmetry, the fundamental field with definite twist is the component
of the gluon field strength tensor~\cite{Braun:2003rp}. The
component $F^{+\perp}$ is twist-1, $F^{\perp\perp}$ and $F^{+-}$ are
twist-2 while $F^{-\perp}$ is twist-3. In the following, we only
consider the leading twist light-cone distribution amplitudes of the
glueball state:
\begin{eqnarray}
 \langle G(P)|F^{a,+\mu}(z^-)F^{b,+\nu}(0)|0\rangle &=& \int_0^1 dx
 e^{ixz^- P^+} P^{+2}\frac{f_s\delta^{ab}}{2(N_c^2-1)}\left[g^{\mu\nu}_\perp \phi_G(x)
 \right],\label{eq:definition-gauge-invariant-LCDA}
\end{eqnarray}
where $\mu,\nu$ are the transverse indices. The coordinate $z$ has
been chosen on the light-cone $z^2=0$. In
Eq.~(\ref{eq:definition-gauge-invariant-LCDA}) we have used the
light-cone gauge so that the gauge links between the field
strength tensors vanish. The distribution amplitude $\phi(x)$ can
be expanded in terms of the Gegenbauer polynomials:
\begin{eqnarray}
 \phi_G(x)&=&
 30x^2(1-x)^2\left[1+\sum_{n}a_nC_n^{5/2}(2x-1)\right].\label{eq:LCDAgegenbauermoments}
\end{eqnarray}
In the $A^+=0$ gauge, the light cone distribution amplitude is
reformulated as:
\begin{eqnarray}
  \langle
 G(P)|A^{a\mu}(z^-)A^{b\nu}(0)|0\rangle &=&- \int_0^1 dx e^{ixz^- P^+}\frac{f_s\delta^{ab}}{2(N_c^2-1)}\left[
 g^{\mu\nu}_\perp \frac{
 \phi_G(x)}{x(1-x)} \right],\label{eq:matrix-element-two-gluons}
\end{eqnarray}
where only the transverse gluon contributes at the leading twist.

Before presenting the formulas for the transition form factors, we
will briefly comment on the transverse distribution of the wave
functions. The basic idea of the PQCD approach is that it takes into
account the transverse momentum of the valence quarks in hadrons
which kills the endpoint singularity. The form factor is expressed
as a convolution of the wave functions and a hard kernel.
Resummation of the double logarithm due to higher order corrections
results in the Sudakov factor.  Strictly speaking the transverse
distribution in all wave functions should be taken into account in
the PQCD approach. One of the most acceptable candidates would be
the exponential wave function or improved exponential form like the
Gaussian one for the $B$ meson given in Eq.~(\ref{Bwave:da}). The
common feature of these distributions is that the contribution from
the large $b$ region is exponentially suppressed. Meanwhile the
Sudakov factor can also suppress the contribution from the large $b$
region as shown in Fig.~2 in the first paper of
Ref.~\cite{Lu:2000em}. This suppression effect also depends on the
longitudinal momentum. Since the momentum of the quark and antiquark
in a light meson is large, contributions from the large $b$ region
are strongly suppressed by the Sudakov factor. As a consequence the
role of the transverse distribution in the wave function for the
final mesons has already been fulfilled and it can be neglected in
the PQCD approach. On the contrary, the Sudakov suppression for the
$B$ meson is not so manifest that the transverse distribution is
required. The commonly-used form in the PQCD approach is the one
given in Eq.~(\ref{Bwave:da}) which will be adopted in this work.

The form factors in the PQCD approach are obtained by evaluating the
three diagrams in Fig.~\ref{diagram:FeynBtof0}. They can be
expressed as the convolution over the longitudinal momentum fraction
and the transverse space coordinates:
 \begin{eqnarray}
 F_1(q^2)&=&4\frac{f_s\sqrt{2N_c}}{N_c^2-1} \pi C_F m_B^2 \int_0^1
        dx_1dx_2\int_0^{\infty}b_1db_1 \phi_B(x_1,b_1)\frac{\phi_G(x_2)}{x_2(1-x_2)}\nonumber\\
        &&\times\left\{ \int_0^{\infty}b_2db_2\left\{x_1\left[\rho+(1-\rho)x_2 \right] E_a(t_a)
        h_a
        +  \rho  x_2  E_c(t_c) h_c\right\}+\rho x_2(1-x_2)  E_b(t_b) h_b \right\},\label{eq:F1}\\
 F_0(q^2)&=&4\frac{f_s\sqrt{2N_c}}{N_c^2-1} \pi C_F m_B^2 \int_0^1
        dx_1dx_2\int_0^{\infty}b_1db_1 \phi_B(x_1,b_1)\frac{\phi_G(x_2)}{x_2(1-x_2)}\nonumber\\
        &&\times\left\{ \int_0^{\infty}b_2db_2\left\{x_1\rho\left[2-\rho+(\rho-1)x_2 \right] E_a(t_a) h_a
        +\rho (2-\rho)x_2 E_c(t_c) h_c\right\}\right.\nonumber\\
        &&\;\;\;\;\left.
        +  \rho^2 x_2(1-x_2) E_b(t_b) h_b \right\},\label{eq:F0}\\
 F_T(q^2)&=& 4\frac{(m_B+m_S)f_s\sqrt{2N_c}}{N_c^2-1} \pi C_F m_B \int_0^1
        dx_1dx_2\int_0^{\infty}b_1db_1 \phi_B(x_1,b_1)\frac{\phi_G(x_2)}{x_2(1-x_2)}\nonumber\\
        &&\times\left\{ \int_0^{\infty}b_2db_2\left\{x_1x_2 E_a(t_a) h_a
        +x_1(1-x_2)E_c(t_c) h_c\right\} -x_1  E_b(t_b) h_b \right\},\label{eq:FT}
 \end{eqnarray}
where  $C_F=4/3$ and $N_c=3$ are color factors. $E_i,h_i$ are hard
functions determined through the propagators in the three diagrams,
which   are collected in the appendix. Through fitting the results
in the hard-scattering region $0<q^2<10\mbox{GeV}^2$, we extrapolate
them with the dipole model parametrization
 \begin{equation}
 F_i(q^2)=\frac{F_i(0)}{1-a(q^2/m_B^2)+b(q^2/m_B^2)^2},\label{eq:fitting-form}
 \end{equation}
where $i=1,0,T$ and $a,b$ are parameters to be determined in the
fitting procedure.

In the numerical calculation, we adopt
$\omega_B=(0.40\pm0.05)\mbox{GeV}$ and $f_B=(0.19\pm0.02)\rm{GeV}$
for $B$ mesons \cite{Keum:2000ph}. For the $B_s$ meson, we use
$\omega_{B_s}=(0.50\pm0.05)\mbox{GeV}$ and
$f_{B_s}=(0.23\pm0.02)\rm{GeV}$ \cite{bs}. The decay constant of the
scalar glueball is used as $f_s=0.13 $GeV~\cite{He:2002hr}, but so
far there is no theoretical study on the LCDAs of the glueball. The
simplest choice is to use the asymptotic form. To roughly estimate
the uncertainty from the higher Gegenbauer moment, we will also try
the Gegenbauer moment $ a_2=0.2. $ The results in
table~\ref{tab:formfactor} show that the form factors are sensitive
to the Gegenbauer moment $a_2$, which will provide relatively
important uncertainties to our results. The mass of the glueball is
taken as $m_G=1.5$ GeV. In the PQCD approach, the $B$-to-glueball
transition form factors are not very sensitive to this mass.



\begin{table}
\caption{Distinct contributions to $B\to G$ form factors at $q^2=0$:
the index (a,b,c) denotes the contribution from the diagram (a,b,c)
in Fig.~\ref{diagram:FeynBtof0}, respectively. Except the results in
the last column where $a_2=0$ is used, all other results are
obtained with $a_2=0.2$. }\label{tab:formfactor}
\begin{tabular}{c|c|c|c|c|c}
 \hline\hline & $a$ & $b$& $c$ & total  & asymptotic \\ \hline
  $B\to G$\ \ \   $F_1=F_0$  &0.042 & 0.012 & 0.001 & 0.055 &0.043 \\ \hline
    \ \ \ \ \ $F_T$  &0.035 &$-0.011$ & 0.001 &0.025& $0.017$  \\
 \hline\hline
  $B_s\to G$\ \ \   $F_1=F_0$  &0.038 & 0.011 & 0.002 & 0.051 &0.039 \\ \hline
    \ \ \ \ \ $F_T$  &0.032 &$-0.012$ & 0.001 &0.020& $0.014$  \\
    \hline\hline
\end{tabular}
\end{table}

 \begin{table}
 \caption{$B\to G$ and $B_s\to G$ form factor with the dipole parametrization.
Results in the "asymptotic" ("total") rows correspond to the
Gegenbauer moment $a_2=0$
($a_2=0.2$).}\label{tab:formfactor-q2-dependence}
 \begin{tabular}{|c|c c c c c c c c|}
 \hline
 \ \ \        &$F_0(0)=F_1(0)$  &$F_T(0)$   &$a(F_1)$    &$b(F_1)$ &$a(F_0)$  &$b(F_0)$    &$a(F_T)$ &$b(F_T)$  \\
 \hline
 $B\to G$\ \ \ asymptotic   &$0.043^{+0.008+0.004}_{-0.007-0.004}$   &$0.017^{+0.004+0.001}_{-0.004-0.001}$  &$1.22^{+0.00}_{-0.02}$  &$0.15^{+0.00}_{-0.02}$  &$0.86^{+0.01}_{-0.02}$  &$0.06^{+0.02}_{-0.03}$  &$1.69^{+0.01}_{-0.03}$  &$0.69^{+0.01}_{-0.06}$\\
 \ \ \ \ total        &$0.055^{+0.011+0.005}_{-0.009-0.004}$   &$0.025^{+0.006+0.002}_{-0.005-0.002}$  &$1.24^{+0.01}_{-0.02}$  &$0.17^{+0.01}_{-0.02}$  &$0.81^{+0.00}_{-0.02}$  &$0.03^{+0.01}_{-0.04}$  &$1.62^{+0.01}_{-0.04}$ &$0.62^{+0.02}_{-0.08}$\\
 \hline
 $B_s\to G$\ \ \ asymptotic   &$0.039^{+0.006+0.004}_{-0.005-0.003}$   &$0.014^{+0.003+0.001}_{-0.003-0.001}$  &$1.20^{+0.01}_{-0.01}$  &$0.13^{+0.02}_{-0.01}$  &$0.84^{+0.02}_{-0.01}$  &$0.04^{+0.03}_{-0.00}$  &$1.64^{+0.02}_{-0.02}$  &$0.63^{+0.05}_{-0.02}$\\
 \ \ \ \ total        &$0.051^{+0.008+0.005}_{-0.007-0.004}$   &$0.020^{+0.005+0.002}_{-0.004-0.002}$  &$1.22^{+0.01}_{-0.01}$  &$0.15^{+0.02}_{-0.01}$  &$0.79^{+0.02}_{-0.01}$  &$0.01^{+0.03}_{-0.01}$  &$1.57^{+0.02}_{-0.02}$  &$0.54^{+0.05}_{-0.01}$\\
 \hline\hline
 \end{tabular}
 \end{table}

 \begin{table}
 \caption{$B\to f_0(1370)$ and $B_s\to f_0(1500)$ form factors in two
 scenarios (denoted by S1 and S2) with the dipole parametrization, where
 $f_0(1370)$ is assigned as $\bar nn$ and $f_0(1500)$ is identified as $\bar ss$.}\label{tab:BtoS-formfactor-s1-s2}
 \begin{tabular}{|c|c c c c c c c c|}
 \hline\hline
 \ \ \         &$F_0(0)$  &$F_T(0)$    &$a(F_1)$    &$b(F_1)$    &$a(F_0)$  &$b(F_0)$         &$a(F_T)$ &$b(F_T)$  \\
 \hline
 $B\to f_0$\ \ \ S1     &$-0.30_{-0.09}^{+0.08}$   &$-0.39_{-0.11}^{+0.10}$  &$1.63_{-0.05}^{+0.09}$  &$0.53_{-0.08}^{+0.14}$  &$0.70_{-0.02}^{+0.07}$  &$-0.24_{-0.05}^{+0.15}$   &$1.60_{-0.04}^{+0.06}$  &$0.50_{-0.05}^{+0.08}$\\
 \ \ \ \ \ \  \ \ \ \ \ \ \ S2     &$0.63_{-0.14}^{+0.23}$   &$0.76_{-0.17}^{+0.37}$    &$1.60_{-0.05}^{+0.15}$  &$0.53_{-0.09}^{+0.18}$  &$0.70_{-0.11}^{+0.05}$  &$-0.14_{-0.09}^{+0.02}$ &$1.63_{-0.05}^{+0.07}$  &$0.57_{-0.07}^{+0.07}$\\
\hline $B_s\to f_0$\ \ \ S1  &$-0.26_{-0.08}^{+0.09}$  &$-0.34_{-0.10}^{+0.10}$  &$0.72_{-0.08}^{+0.14}$  &$-0.20_{-0.10}^{+0.10}$  &$1.61_{-0.03}^{+0.13}$  &$0.48_{-0.02}^{+0.27}$  &$1.60_{-0.04}^{+0.06}$   &$0.48_{-0.04}^{+0.09}$\\
 \ \ \ \ \ \  \ \ \ \ \ \ \ S2    &$0.60_{-0.12}^{+0.20}$  &$0.82_{-0.16}^{+0.30}$  &$0.65_{-0.10}^{+0.04}$  &$-0.22_{-0.02}^{+0.07}$  &$1.76_{-0.08}^{+0.13}$  &$0.71_{-0.08}^{+0.20}$  &$1.71_{-0.07}^{+0.04}$   &$0.66_{-0.10}^{+0.06}$\\
 \hline\hline
 \end{tabular}
 \end{table}

The distinct contributions from the three diagrams are collected in
table~\ref{tab:formfactor}. Among them, the largest contribution is
from the first diagram, while the third one is smallest. This is in
agreement with the power counting rule given in
Sec.~\ref{sec:powercounting}. Although the third diagram can be
neglected compared with other potential contributions like the
higher order corrections and higher power corrections, this diagram
is taken into account since it has the same order in $\alpha_s$ with
the other diagrams. Compared with the gluonic contributions to the
$B\to \eta$ form factors in Ref.~\cite{Charng:2006zj}, our results
are much larger. For the pseudoscalar meson $\eta$ and $\eta'$, the
gluonic LCDAs are normalized to 0, thus only higher Gegenbauer
moments contribute. The Gegenbauer moment used for the gluonic
content of $\eta$ and $\eta'$ is very small $B_2=\frac{2.1}{30}\sim
\frac{7.1}{30}$. Moreover, we have taken into account the second
diagram, which also makes the form factors larger.

In the following, we will comment on the magnitude of
$B$-to-glueball form factors and compare with other transition form
factors.  In the PQCD approach, one ingredient for the form factor
is the matrix element of the gluon fields and the quark fields.
These matrix elements differ with the LCDAs by some constants which
can be explicitly given in the definition of the LCDAs. For the
glueball state the matrix element associated with the asymptotic
twist-2 LCDA defined in Eq.~(\ref{eq:matrix-element-two-gluons}) can
be rewritten as
\begin{eqnarray}
\frac{1}{2(N_c^2-1)}\frac{\phi_G(x)}{x(1-x)}=
\frac{30x(1-x)}{2(N_c^2-1)},\nonumber
\end{eqnarray}
where we have substituted the asymptotic form for the LCDA
$\phi_G(x)$. Similarly, the matrix element defined by the quark
fields for a pseudoscalar meosn is given by
\begin{eqnarray}
 \frac{\phi_P(x)}{4N_c}= \frac{6x(1-x)}{4N_c},\nonumber
\end{eqnarray}
where the asymptotic form for the twist-2 LCDA $\phi_P(x)$ is used:
$\phi_P(x)=6x(1-x)$~\cite{Braun:2003rp}.  The prefactor for the
glueball is almost 4 times larger than that for a pseudoscalar
meson. Despite it, the $B$-to-glueball form factors are smaller than
the $B\to \pi$ form factors. Compared with our previous
studies~\cite{Wang:2006ria,Li:2008tk} on the transition form factors
of $B$ meson decays into ordinary scalar mesons in
Tab.~\ref{tab:BtoS-formfactor-s1-s2}, the $B\to G$ form factors are
smaller by a factor of $6-10$. One reason is that the decay constant
for glueball is smaller than the scalar decay constant for the quark
content. For example, the scalar decay constant for $a_0(1450)$ is
$-0.28\pm0.03$ or $0.46\pm0.05$~\cite{Cheng:2005nb}. There exist
some other reasons. In the transition form factors for the quark
content, the small momentum fraction $x_1$ in the numerator of the
first diagram in Fig.~\ref{diagram:FeynBtof0} is replaced by the
factor accompanied with the twist-3 LCDA: $2r_S=\frac{2m_S}{m_B}$.
Although all of these two terms have the same power, the transition
for the quark content is much larger than that for the gluon
component since $2r_S\gg \langle x_1\rangle$ ($ \langle x_1\rangle $
is the typical value of $x_1$). For the second diagram, the
longitudinal momentum of the quark propagators are given as:
$x_2\bar x_2x_1$, while the corresponding one that appears in the
quark transition form factor is $\bar x_2^2 x_1$. In the latter
case, the region with small values of $\bar x_2$ will give
relatively large contributions although the endpoint singularities
are removed by the transverse momentum and the threshold
resummation. In the former transition, the form factor is directly
reduced by the propagators and the contributions from the small
momentum fraction region are small. The third diagram is power
suppressed as we have discussed. Thus the total form factors are
smaller than those for quark contents.

\begin{figure}
\begin{center}
\includegraphics[scale=0.9]{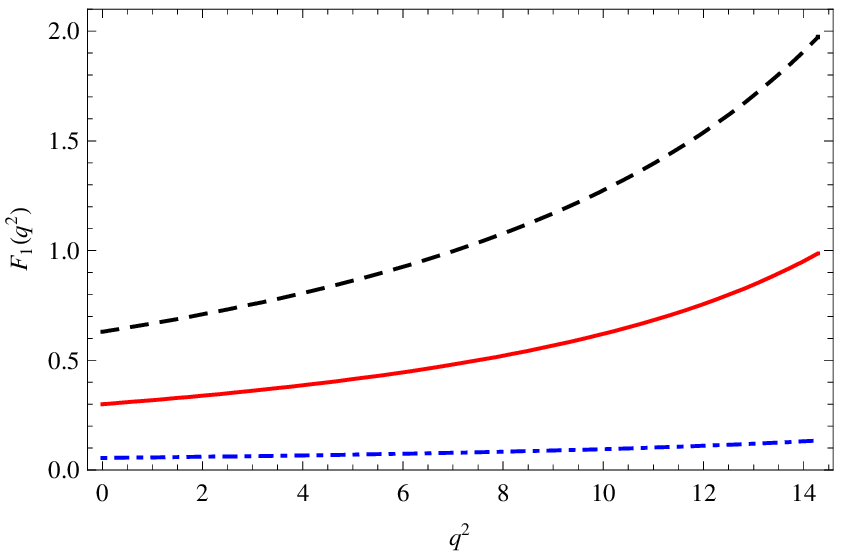}\hspace{1cm}
\includegraphics[scale=0.9]{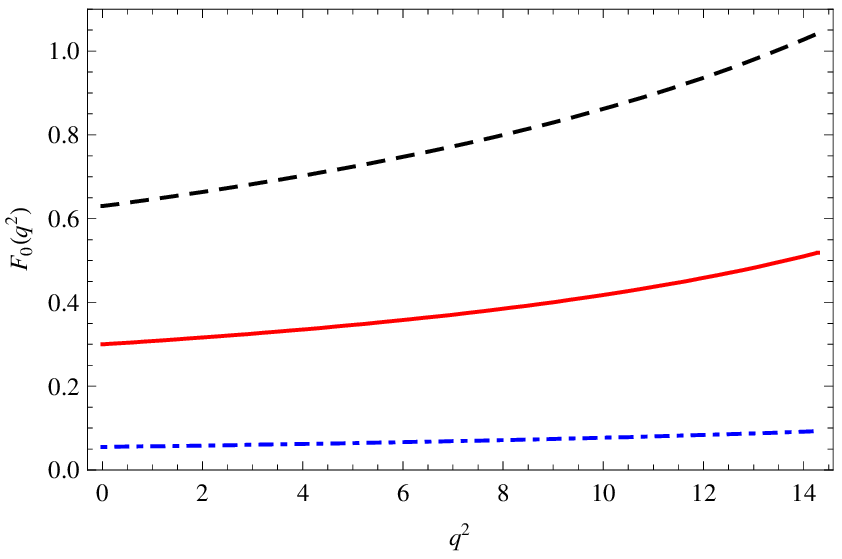}
\end{center}
\caption{ The $q^2$-dependence of $B \to S$ form factors
$|F_1(q^2)|$ and $|F_0(q^2)|$, where the dashed (black) and solid
(red)lines denote the form factors for ordinary mesons in scenario I
and scenario II, while the dash-dotted (blue) lines denote the $B\to
G$ form factors. } \label{diagram:q2dependence}
\end{figure}

Results for the $B\to G$ form factors, together with the
$q^2$-dependent parameters in Eq.~\eqref{eq:fitting-form},  are
collected in Tab.~\ref{tab:formfactor-q2-dependence}. The
uncertainties are from the input parameters: (i) the $B$ meson decay
constant and the shape parameter $(\omega_B,\omega_{B_s})$ in the
wave functions; (ii) the factorization scale from $0.75t$ to $1.25t$
(not changing the transverse part $1/b_i$):
\begin{eqnarray}
 {\rm max}\left[0.75\sqrt {\rho x_1}m_B,1/b_1,1/b_2\right]< t_a<
 {\rm max}\left[1.25\sqrt {\rho x_1}m_B,1/b_1,1/b_2\right],\nonumber\\
  {\rm max}\left[0.75\sqrt {\rho x_2}m_B,\sqrt{\rho\bar x_2
  x_1}m_B,1/b_1\right]<t_b<
  {\rm max}\left[1.25\sqrt {\rho x_2}m_B,\sqrt{\rho\bar x_2 x_1}m_B,1/b_1\right],\nonumber\\
 {\rm max}\left[0.75\sqrt {\rho }m_B,1/b_1,1/b_2\right]<t_c<{\rm max}\left[1.25\sqrt {\rho }m_B,1/b_1,1/b_2\right],
\nonumber
\end{eqnarray} and the hadronic scale $\Lambda_{\rm QCD}=(0.25\pm0.05)$ GeV.
The $q^2$-dependence of the $B\to G$ form factors, together with the
form factors for ordinary scalar mesons, are given in
Fig.~\ref{diagram:q2dependence}. The dashed (black) and solid (red)
lines denote the form factors for ordinary mesons in scenario I and
scenario II, while the dash-dotted (blue) lines denote the $B\to G$
form factors.

Although form factors of $B$ decays into scalar glueballs are
smaller than those for the quark content, one can not neglect the
gluon content in the case of mixing for scalar mesons. There is a
nontrivial factor $1/\sqrt 2$ in form factors for the quark content
in $B$ meson decays. This factor will make the glue component in a
scalar meson more important. Compared with the ordinary $B\to S$
form factors, the PQCD calculation of the $B\to G$ form factors is
expected to be more reliable. In the ordinary $B\to S$ form factors,
the perturbative hard-scattering diagrams contain the endpoint
singularity in the collinear factorization. Although the inclusion
of the transverse momentum can resolve this problem, the results
would still be sensitive to the treatment of the endpoint region. In
the PQCD approach resummation of the double logarithms results in
the Sudakov factor which will suppress contributions from the
nonperturbative region (large $b$ region). This will improve the
convergence of the perturbation theory.  These Sudakov effects in
$B\to \pi$ transition have also been investigated in
Ref.~\cite{DescotesGenon:2001hm} with a different conclusion (see
Ref.~\cite{Li:2004ja} for the response). On the contrary, the
situation is different in the $B$-to-glueball transition which is
free from the endpoint singularity. The results are more stable than
the $B\to S$ form factors.

\section{Phenomenological Applications}
\label{eq:numericalresults}
\subsection{Semileptonic $B\to Gl\bar\nu$ and $B\to Gl^+l^-$ decays}
%
%

Integrating out the highly off-shell intermediate degrees of
freedom, the effective Hamiltonian for $b\to ul\bar \nu_l$
transition is given by~\cite{Buchalla:1995vs}
 \begin{eqnarray}
 {\cal H}_{\rm eff}(b\to ul\bar \nu_l)=\frac{G_F}{\sqrt{2}}V_{ub}\bar
 u\gamma_{\mu}(1-\gamma_5)b \bar l\gamma^{\mu}(1-\gamma_5)\nu_l.
 \end{eqnarray}
while the effective Hamiltonian responsible for the  $b\to Dl^+l^-$
$(D=d,s)$ transition is
 \begin{eqnarray}
 {\cal
 H}_{\rm{eff}}=-\frac{G_F}{\sqrt{2}}V_{tb}V^*_{tD}\sum_{i=1}^{10}C_i(\mu)O_i(\mu),\label{eq:Hamiltonian}
 \end{eqnarray}
where $V_{ub},V_{tb}$ and $V_{tD}$ $(D=d,s)$ are the
Cabibbo-Kobayashi-Maskawa (CKM) matrix elements. $C_i(\mu)$ are the
Wilson coefficients and the local operators $O_i(\mu)$ are given by
 \begin{eqnarray}
 O_1&=&(\bar D_{\alpha}c_{\alpha})_{V-A}(\bar
 c_{\beta}b_{\beta})_{V-A},\;\;
 O_2=(\bar
 D_{\alpha}c_{\beta})_{V-A}(\bar
 c_{\beta}b_{\alpha})_{V-A},\nonumber\\
 O_3&=&(\bar D_{\alpha}b_{\alpha})_{V-A}\sum_q(\bar
 q_{\beta}q_{\beta})_{V-A},\;\;
 O_4=(\bar D_{\alpha}b_{\beta})_{V-A}\sum_q(\bar
 q_{\beta}q_{\alpha})_{V-A},\nonumber\\
 O_5&=&(\bar D_{\alpha}b_{\alpha})_{V-A}\sum_q(\bar
 q_{\beta}q_{\beta})_{V+A},\;\;
 O_6=(\bar D_{\alpha}b_{\beta})_{V-A}\sum_q(\bar
 q_{\beta}q_{\alpha})_{V+A},\nonumber\\
 O_7&=&\frac{e m_b}{8\pi^2}\bar
 D\sigma^{\mu\nu}(1+\gamma_5)bF_{\mu\nu}+\frac{e m_D}{8\pi^2}\bar
 D\sigma^{\mu\nu}(1-\gamma_5)bF_{\mu\nu},\nonumber\\
 O_9&=&\frac{\alpha_{\rm{em}}}{2\pi}(\bar l\gamma_{\mu}l)(\bar
 D\gamma^{\mu}(1-\gamma_5)b),\;\;
 O_{10}=\frac{\alpha_{\rm{em}}}{2\pi}(\bar l\gamma_{\mu}\gamma_5l)(\bar
 DZ\gamma^{\mu}(1-\gamma_5)b),\label{eq:operators}
 \end{eqnarray}
where $(\bar q_1q_2)_{V-A}(\bar q_3 q_4)_{V-A}\equiv(\bar q_1
\gamma^{\mu}(1-\gamma_5)q_2)(\bar q_3\gamma_{\mu}(1-\gamma)q_4)$,
and $(\bar q_1q_2)_{V-A}(\bar q_3 q_4)_{V+A}\equiv(\bar q_1
\gamma^{\mu}(1-\gamma_5)q_2)(\bar q_3\gamma_{\mu}(1+\gamma)q_4)$.
$\alpha,\beta$ are the color indices for the quark field.
$\alpha_{\rm em}=1/137$ is the fine structure constant. The
amplitude for $b\to sl^+l^-$ transition can be decomposed as
 \begin{eqnarray}
 {\cal A}(b\to sl^+
 l^-)&=&\frac{G_F}{2\sqrt{2}}\frac{\alpha_{\rm{em}}}{\pi}V_{tb}V^*_{ts}\bigg\{
 C_9^{\rm{eff}}(q^2)
 [\bar s \gamma_{\mu}(1-\gamma_5)b][\bar l\gamma^{\mu}l] + C_{10}[\bar s\gamma_{\mu}(1-\gamma_5)b]
 [\bar l\gamma^{\mu}\gamma_5l]\nonumber\\
 &&- 2m_bC_7^{\rm{eff}}\big[\bar s i\sigma_{\mu\nu}\frac{q^{\nu}}{q^2}
 (1+\gamma_5)b\big][\bar l\gamma^{\mu}l]- 2m_sC_7^{\rm{eff}}\big[\bar s i\sigma_{\mu\nu}\frac{q^{\nu}}{q^2}
 (1-\gamma_5)b\big ][\bar l\gamma^{\mu}l] \bigg\},\label{eq:Ampbtos}
 \end{eqnarray}
with $m_b$ as the $b$ quark mass in the $\overline{\mbox{MS}}$
scheme and we will use $m_b=4.8$ GeV. $C_7^{\rm{eff}}=C_7-C_5/3-C_6$
and $C_9^{\rm{eff}}$ contains both the long-distance and
short-distance contributions, which is given by
 \begin{eqnarray}
 C_9^{\rm{eff}}(q^2)&=&C_9(\mu)+Y_{\rm{pert}}(\hat{s})+Y_{\rm{LD}}(q^2).
 \end{eqnarray}
with $\hat{s}=q^2/m_B^2$. $Y_{\rm{pert}}$ represents the
perturbative contributions, and $Y_{\rm{LD}}$ is the long-distance
part. Since the long-distance contribution can be separated on the
experimental side, we will neglect it.  The short-distance
corrections $Y_{\rm{pert}}$ is given by~\cite{Buras:1994dj}
 \begin{eqnarray}
 Y_{\rm{pert}}(\hat{s})&=&
 h(\hat{m_c},\hat{s})C_0-\frac{1}{2}h(1,\hat{s})(4C_3+4
 C_4+3C_5+C_6)\nonumber\\
 &&-\frac{1}{2}h(0,\hat{s})(C_3+3
 C_4) + \frac{2}{9}(3C_3 + C_4 +3C_5+ C_6),\label{eq:ypert}
 \end{eqnarray}
with $C_0=C_1+3C_2+3C_3+C_4+3C_5 +C_6$ and $\hat{m}_c=m_c/m_b$ with
$m_c=1.27$ GeV~\cite{Amsler:2008zzb}. The relevant Wilson
coefficients, listed in table \ref{tab:wilsons}, are given up to the
leading logarithmic accuracy~\cite{Buchalla:1995vs}.

\begin{table}
 \caption{The values of Wilson coefficients $C_i(m_b)$ in the leading
logarithmic approximation, with $m_W=80.4\mbox{GeV}$, $\mu=m_{b,\rm
pole}$~\cite{Buchalla:1995vs}.}
 \label{tab:wilsons}
 \begin{center}
 \begin{tabular}{c c c c c c c c c}
 \hline\hline
 \ \ \ $C_1$ &$C_2$ &$C_3$ &$C_4$ &$C_5$ &$C_6$ &$C_7^{\rm{eff}}$ &$C_9$ &$C_{10}$       \\
 \ \ \ $1.107$   &$-0.248$   &$-0.011$    &$-0.026$    &$-0.007$    &$-0.031$    &$-0.313$    &$4.344$    &$-4.669$    \\
 \hline\hline
 \end{tabular}
 \end{center}
 \end{table}

Lepton decay amplitudes can be directly calculated using the
perturbation theory. The unknown amplitude is the matrix elements of
quark operators which have been parameterized as the form factors.
The partial decay width is given by
 \begin{eqnarray}
 \frac{d\Gamma(B\to Gl\bar\nu)}{dq^2}&=&\frac{\sqrt \lambda G_F^2|V_{ub}|^2}{384 \pi^3
 m_B^3q^2}\left(\frac{q^2-m_l^2}{q^2}\right)^2\bigg[(m_l^2+2q^2)\lambda F_1^2(q^2)
 +3m_l^2(m_B^2-m_S^2)^2F_0^2(q^2)\bigg],\label{eq:BtoGlnudgammadqsquare}
 \end{eqnarray}
where $\lambda=(m_B^2-q^2-m_G^2)^2-4m_G^2q^2$ and $m_l$ is the
lepton mass. Integrating out the momentum transfer square $q^2$, we
obtain branching ratios of $B\to Gl\bar\nu$ ($l=e,\mu$) and $B\to
G\tau\bar\nu_\tau$
\begin{eqnarray}
 {\cal B}(B\to Gl\bar\nu)&=&(0.18^{+0.08+0.03+0.02}_{-0.06-0.03-0.02} [0.11^{+0.05+0.02+0.01}_{-0.03-0.02-0.01}])\times 10^{-5},\label{17}\\
 {\cal B}(B\to G\tau\bar\nu_\tau)&=&(0.08^{+0.03+0.01+0.01}_{-0.03-0.01-0.01}[0.05^{+0.02+0.01+0.00}_{-0.02-0.01-0.00}])\times
 10^{-5},\label{18}
\end{eqnarray}
where results in the square brackets and out of the square
brackets are evaluated using the asymptotic form and the form with
$a_2=0.2$, respectively. In the evaluation of semileptonic $B$
decays we have adopted the dipole form for the form factors given
in Eq.~\eqref{eq:fitting-form}, where the parameters are obtained
through fitting the large recoil region. Uncertainties are from
three different kinds of inputs: (i) the $B$ meson decay constant
and the shape parameter $(\omega_B,\omega_{B_s})$, (ii) the
hadronic scale $\Lambda_{\rm QCD}$, the factorization scale $t$,
(iii) the CKM matrix element
$|V_{ub}|=(3.51^{+0.14}_{-0.16})\times
10^{-3}$~\cite{Charles:2004jd}. Compared with the recently
measured result of $B\to \eta l\bar\nu$ decay~\cite{Aubert:2008ct}
\begin{eqnarray}
 {\cal B}(B^-\to \eta l^-\bar\nu)&=&(3.1\pm0.6\pm0.8)\times
 10^{-5},
\end{eqnarray}
we can see that $B\to Gl\bar\nu (l=e,\mu)$ have smaller branching
ratios by one order. The main decay channel of a scalar glueball
could be $\pi\pi$ or $K\bar K$ while the $\eta$ meson is
reconstructed by three pions or two photons. Thus a scalar glueball
is easier to detect than the isoscalar meson $\eta$, and the $B\to
Gl\bar\nu$ decays may be observed in the future.

The partial decay width of semileptonic  $B_{d,s}\to Gl^+l^-$ decays
is given as
 \begin{eqnarray}
 \frac{d\Gamma(B_{d,s}\to
Gl^+l^-)}{dq^2}
 &=&\sqrt {\frac{q^2-4m_l^2}{q^2}}\frac{G_F^2 \alpha_{\rm em}^2  \sqrt \lambda }{1024m_B^3\pi^5}\times
 |V_{tb}V_{tD}|^2
 \nonumber\\
 &&\times \left[\frac{4}{3} \lambda  \left| \frac{C_9^{\rm eff}}{2}
   F_1(q^2)
   +\frac{ C_{10}}{2}   F_1(q^2) \sqrt {\frac{q^2-4m_l^2}{q^2}}   +(C_{7L}-C_{7R})
  \frac{m_b F_T(q^2)}{ m_B+m_P }\right|^2\right. \nonumber\\
 &&\left.+\frac{4}{3} \lambda  \left| \frac{C_9^{\rm eff} }{2}
   F_1(q^2)
   -
  \frac{ C_{10}}{2}   F_1(q^2) \sqrt {\frac{q^2-4m_l^2}{q^2}}  +(C_{7L}-C_{7R})
  \frac{m_b F_T(q^2)}{(m_B+m_P)} \right|^2\right.\nonumber\\
  &&\left. +\frac{4\lambda m_l^2}{3q^2}\left| C_9^{\rm eff}
  F_1(q^2)  +(C_{7L}-C_{7R})
  \frac{2 m_bF_T(q^2) }{ (m_B+m_P)}
 \right|^2 +\frac{4m_l^2}{q^2}\left|
 C_{10}(m_B^2-m_P^2)F_0(q^2)\right|^2\right],
 \end{eqnarray}
where $D=d,s$ for $\bar B^0\to Gl^+l^-$ and $\bar B_s^0\to Gl^+l^-$
decays.  $m_l$ denotes the lepton's mass and $C_{7L}=C_7^{\rm eff}$,
$C_{7R}=\frac{m_{s,d}}{m_b}C_7^{\rm eff}\sim 0$. Branching fractions
of $B\to Gl^+l^-$ and $B_s\to Gl^+l^-$ decays are predicted as
\begin{eqnarray}
  {\cal B}(B\to G l^+l^-)&=&(0.27^{+0.12+0.05}_{-0.09-0.04}[0.18^{+0.07+0.03}_{-0.05-0.03}])\times 10^{-9},\label{20}\\
  {\cal B}(B\to G \tau^+\tau^-)&=&(0.5^{+0.2+0.1}_{-0.2-0.1}[0.3^{+0.1+0.1}_{-0.1-0.1}])\times 10^{-11},\\
  {\cal B}(B_s\to G l^+l^-)&=&(0.6^{+0.2+0.1}_{-0.2-0.1}[0.4^{+0.1+0.1}_{-0.1-0.1}])\times 10^{-8},\\
  {\cal B}(B_s\to G \tau^+\tau^-)&=&(0.18^{+0.07+0.04}_{-0.05-0.03}[0.12^{+0.04+0.02}_{-0.03-0.02}])\times
  10^{-9}.\label{23}
\end{eqnarray}
Uncertainties are from two different kinds of inputs in the PQCD
approach: (i) the $B$ meson decay constant and the shape parameter
$(\omega_B,\omega_{B_s})$, (ii) the hadronic scale $\Lambda_{\rm
QCD}$, the factorization scale $t$. The CKM matrix element
$V_{td}=(8.59^{+0.28}_{-0.29})\times 10^{-3}$,
$V_{ts}=-(40.42^{+1.18}_{-0.37})\times 10^{-3}$ and
$V_{tb}=0.999146^{+0.000048}_{-0.000016}$~\cite{Charles:2004jd} will
not give large uncertainties. The $B\to Gl^+l^-$ has tiny branching
fractions, which can not be observed on the present $B$ factories,
especially the $B\to G\tau^+\tau^-$ with a very small phase space.
But the $B_s\to Gl^+l^-$ decay channel may be observed on the future
experiment, since it has a sizable branching fraction.

The physical scalar meson is probably a mixture of glueball and
ordinary states, so the predicted branching ratios in
eq.(\ref{17},\ref{18},\ref{20}-\ref{23}) will be  smaller for
physical scalar meson, according to the gluon component of each
scalar meson. The experimental accessability will still be large, at
least for the scalar mesons with large glue components.

%

\subsection{Mixing between scalar mesons}

There is not any solid experimental evidence for a pure glueball
state up to now. Lattice QCD predicted the mass of a scalar glueball
ground state around $1.5$-1.8 GeV~\cite{Bali:1993fb,Chen:2005mg}. At
present most of the lattice studies on hadronic correlators are in
the quenched approximation by neglecting the fluctuations of the
quarks. Due to the lack of dynamical quarks, the binding of
quark-antiquark systems must be attributed to the nonperturbative
properties of gluons, the unique dynamical degree of freedom in the
Lattice QCD. Secondly the simulations are based on the computation
of the matrix element of the gluon operators, while the glueball
states $G$ are obtained by smeared gluonic operators since there is
no physical glueball state. Nevertheless, despite these ambiguities
the Lattice QCD simulations give us a hint that one scalar glueball
state might exist around this mass region. It is very likely that
the glueball state mixes with the ordinary quark-antiquark state to
form several physical mesons.

In the literature, three scalar mesons $f_0(1370)$, $f_0(1500)$ and
$f_0(1710)$ are usually considered as the potential candidates. The
mixing matrix can be generally set as
\begin{eqnarray}
 \left( \begin{array}{c}
 f_0(1710) \\f_0(1500)\\ f_0(1370)
 \end{array} \right) = \left( \begin{array}{ccc} a_1 & a_2
&a_3 \\b_1 &b_2 &b_3  \\c_1 &c_2 &c_3\end{array}\right)
\left(\begin{array}{c} G \\ \bar ss \\ \bar nn\end{array}
\right).\label{1}
\end{eqnarray}
The unitary condition implies that the mixing matrix has only
three independent real numbers. A non-zero $a_1$, $b_1$ or $c_1$
would be a clear evidence for the existence of a glueball.   The
semileptonic $B\to f_0l\bar\nu$ decays receive contributions from
the $\bar nn$ component and glue component but without $\bar ss$
component (at least negligible), while the semileptonic $B_s\to
f_0 l^+l^-$ channel only receive contributions from the $\bar ss$
and glue component but without $\bar nn$ component.   Thus the
experimental measurements for the two channels can give effective
constraints to the mixing parameters.   For the three kinds of
flavor singlet scalar mesons, we have altogether 6 experiments,
but only three real parameters in eq.(\ref{1}) to be fixed. Since
the branching fraction of $B_s\to f_0l^+l^-$ is expected to have
the order of $10^{-8}$ or even smaller, one needs to accumulate a
large number of $B$ decay events, which could be achieved on the
future experiments such as the Super B factory.

Taking into account the current available experimental data, the
mixing mechanism is still not fixed, for example there are two quite
different mixing matrix determined by two group of people. Because
the decay width of $f_0(1500)$ is not compatible with the ordinary
$\bar qq$ state, Amsler and Close claimed that $f_0(1500)$ is
primarily a scalar glueball~\cite{Amsler:1995tu}. In the subsequent
studies~\cite{Close:2000yk}, they extracted the mixing matrix
through fitting the data of two-body decays of scalar mesons
\begin{eqnarray}
 \left( \begin{array}{c}
 f_0(1710) \\f_0(1500)\\ f_0(1370)
 \end{array} \right)
 = \left( \begin{array}{ccc} 0.36 & 0.93&0.09 \\
 -0.84 &0.35 &-0.41  \\
 0.40 &-0.07 &-0.91\end{array}\right)
\left(\begin{array}{c} G \\ \bar ss \\ \bar nn\end{array}
\right).\label{2}
\end{eqnarray}
Based on the SU(3) assumption for scalar mesons and the quenched
Lattice QCD results \cite{Chen:2005mg}, Cheng et
al.~\cite{Cheng:2006hu} reanalyze all existing experimental data and
fit the mixing coefficient as
\begin{eqnarray}
 \left( \begin{array}{c}
 f_0(1710) \\f_0(1500)\\ f_0(1370)
 \end{array} \right)
 = \left( \begin{array}{ccc} 0.93 & 0.17&0.32 \\
 -0.03 &0.84 &-0.54  \\
 -0.36 &0.52 &0.78\end{array}\right)
\left(\begin{array}{c} G \\ \bar ss \\ \bar nn\end{array}
\right).\label{3}
\end{eqnarray}
Here, the $f_0(1710)$ tends to be a primary glueball. This is very
different from the first matrix of mixing coefficients in (\ref{2}).
The scalar meson production rates in $B$ meson decays can be used to
distinguish these assignments, starting with the $B\to S$ form
factors   collected in Tab.~\ref{tab:formfactor-q2-dependence} and
Tab.~\ref{tab:BtoS-formfactor-s1-s2}.
\begin{itemize}
\item In scenario I, if we use the mixing
coefficients in Eq.~(\ref{2}), the production rates of $f_0(1710)$
and $f_0(1500)$ in $B$ decays are much smaller than that of
$f_0(1370)$ but they have large and comparable production rates in
$B_s$ decays; if we use the mixing coefficients in Eq.~(\ref{3}),
$f_0(1710)$ has small production rates in both $B$ and $B_s$
decays but the other two mesons have large and comparable
production rates in $B$ and $B_s$ decays.

\item In scenario II, if we use the mixing coefficients in
Eq.~(\ref{2}), production rates of $f_0(1370)$ and $f_0(1500)$ in
$B$ decays are large and comparable, and $f_0(1710)$ is copious in
$B_s$ decays; if we use the mixing coefficients in Eq.~(\ref{3}),
three mesons have similar production rates in $B$ decays, and
$f_0(1500)$ is more copious than $f_0(1370)$ and $f_0(1710)$ in
$B_s$ decays.
\end{itemize}
Based on our predictions on form factors in this work and in our
previous studies~\cite{Wang:2006ria,Li:2008tk}, these differences in
$B$ and $B_s$ decays are helpful to distinguish the two mixing
mechanisms. Once the branching fraction of these decays were
measured, our calculation can be used to constrain the mixing angle.
One more practical method in future would be to use the $\chi^2$-fit
method and take the theoretical and experimental uncertainties into
account. This requires the future experimental studies.

The main decay channel of a scalar meson would be $G\to PP$, and
thus the scalar meson could be reconstructed from $\pi^+\pi^-$ or
$K^+K^-$. The scalar meson may also decay into a pair of vectors and
thus it could be reconstructed as 4 pions. For the three candidates
of glueball states, the reconstruction method is different.  If
$f_0(1370)$ is the dominant glueball state, the best candidate is
from 4-pion state, since its main decay mode is
$f_0(1370)\to\rho\rho$~\cite{Amsler:2008zzb}. For $f_0(1500)$, the
dominant mode is $f_0(1500)\to\pi\pi (34.9\pm2.3\%)$ and
$f_0(1500)\to 4\pi(49.5\pm3.3\%)$. As for $f_0(1710)$, the dominant
decay mode is $f_0(1710)\to K\bar K$.

\subsection{Uncertainty analysis}

Besides the uncertainties that we have already given in the above,
there exist several other uncertainties, which may affect the
extraction of the mixing matrix of scalar mesons.

\begin{itemize}

\item Our   results depend on the form factors calculated in the
perturbative QCD approach, which is based on the $k_T$
factorization. The  application of PQCD approach  to many channels
involving the s-wave mesons is successful. Phenomenologically, the
large annihilations can explain the correct branching ratios and
direct CP asymmetries of $\bar B^0\to\pi^+\pi^-$ and $\bar B^0\to
K^-\pi^+$~\cite{Hong:2005wj}, the polarization problem of $B\to \phi
K^*$~\cite{Li:2004mp}, etc. Up to now, this approach is only at the
leading order in $1/m_B$ expansion, without the next-to-leading
order corrections. Since the scalar glueball is not very light, the
power corrections proportional to $m_S/m_B$ may also be important.
Moreover although this approach has been proved for the $B\to \pi$
form factor~\cite{Nagashima:2002ia}, there is no proof for the
factorization of the $B$-to-glueball form factor and this may
introduce some potential uncertainties.

\item The hadronic inputs give another origin of the
uncertainties. The decay constant $f_s$ of a scalar glueball may
introduce sizable uncertainties (roughly $20\%$).   Since there is
not any theoretical study on the LCDA of scalar glueball, the
higher but unknown Gegenbauer moments in the LCDA may also provide
sizable contributions.

\item In the presence of mixing, the uncertainties from ordinary
scalar mesons  also affect the production rates of scalar
glueballs sizably. For example in the two different scenarios for
scalar mesons, the form factors for the quark states even have
different signs, because of the different signs in the decay
constants of ordinary scalar mesons~\cite{Cheng:2005nb}.
Accordingly, the interference between quark content and gluon
content differs in signs in the two scenarios. Moreover, the
magnitudes of decay constants could also suffer from large
ambiguities. We take the decay constant of $f_0(980)$ as an
example. Based on the two-quark assumption for the $f_0(980)$
meson,  the early study in the QCD sum rule~\cite{DeFazio:2001uc}
gives
\begin{eqnarray}
 \bar f_{f_0(980)}&=& (0.18\pm 0.015) {\rm GeV},\nonumber
\end{eqnarray}
while the recent study~\cite{Cheng:2005nb} predicts two sets of
values
\begin{eqnarray}
 \bar f_{f_0(980)}&=& (0.37\pm 0.02){\rm GeV} [(0.46\pm0.025){\rm GeV}],\nonumber
\end{eqnarray}
at the scale $\mu=1$GeV ($\mu=2.1$GeV). The decay constants for
the isosinglet mesons near 1.5 GeV may also suffer from similar
uncertainties and the interference between different components
will be complicated.
\end{itemize}
These quantities will inevitably affect the production rates of
scalar mesons, and the extraction of the mixing matrix will
require more precise, both experimental and theoretical, studies.
For example, the experimental study on semileptonic $B\to
a_0(1450)$ decays is helpful to constrain the contributions from
the quark content. On the theoretical side uncertainties caused by
the hadronic inputs could be reduced in a systematic and
comprehensive study on the ordinary scalar meson and the glueball
state. This is beyond the scope of this work and will be reported
elsewhere.

\subsection{Comparison with other works}

In Ref.~\cite{He:2006qk}, the authors investigated the inclusive
$B\to X_s G$ decay, where the large production rate of $f_0(1710)$
is viewed as the signal of a scalar glueball. The rate of $B$
decays into a scalar glueball through gluonic penguin is expected
to be sizable, but $B\to X_s G$ is an inclusive mode, which is
more difficult than exclusive $B$ meson decays experimentally. The
background from the other mesons may also pollute their method.
The authors in Ref.~\cite{Minkowski:2004xf} used symmetries of the
penguin contributions and studied the production rates of a scalar
glueball in charmless three-body $B$ decays. Irrespective of the
validity of neglecting contributions from tree operators (with
large Wilson coefficients) and the uncertainties from the symmetry
breaking effect, much more data on the $B\to S$ decays is required
to make precise predictions.

In Ref.~\cite{Chen:2007uq}, the author studied  the exclusive
channel $B\to G K(K^*)$ which could be experimentally detected.
Since it is a nonleptonic decay, the uncertainties would be larger
than those in $B\to Gl\bar\nu$ and $B_{d,s}\to Gl^+l^-$ decays.
$B\to GK$ is a penguin-dominated process, which is purely induced
by the loop effect in the standard model. The relevant Wilson
coefficient is small and the result is   sensitive to the
next-to-leading order correction and/or the potential new physics
effect. Hadronic uncertainties are also typically large in
nonleptonic decays. Contributions from the $B\to G$ transition
form factor are neglected in their analysis. The light-cone
distribution amplitude is not used in their work, thus the
internal structure of a scalar glueball could not completely
reflected. It would be interesting to reanalyze this nonleptonic
decay channel by combining our prediction on the form factors and
the other amplitudes studied in their work.

Compared with the studies in the literature, we can see that
albeit the production rates of a glueball state in semileptonic
$B$ decays are not very large, the chosen channels $B\to
Gl\bar\nu$ and $B_s\to Gl^+l^-$ are easily measured and rather
clean. In inclusive decays $B\to X_s G$ decays, it is almost
impossible to study the mixing between ordinary scalar mesons and
a glueball state.

\subsection{Some discussions on   nonleptonic $B$ decays}

Semileptonic $B$ decays are clean but in $B\to f_0l\bar\nu$, the
neutrino is identified as missing energy and the efficiency is
limited; while the $B_s\to f_0l^+l^-$ has a small branching ratio.
In these decays, the lepton pair does not carry any SU(3) flavor
and the decay amplitudes receive less pollution from the strong
interactions. The lepton pair can also be replaced by a charmonium
state such as $J/\psi$ since $J/\psi$ does not carry any light
flavor either. $B\to J/\psi f_0$ decays may provide another ideal
probe to detect the internal structure of the scalar mesons. In
$B\to J/\psi f_0$ decay, the $\bar ss$ component will not
contribute at the leading order in $\alpha_s$. For example, the
$B\to J/\psi\phi$ decay has been set a very stringent upper
limit~\cite{:2008bt}: $ {\cal B}(B\to J/\psi\phi)<9.4\times
10^{-7}$. Thus $B\to J/\psi f_0$ decay can filter out the glueball
component and the $\bar nn$ component of a scalar meson. Meanwhile
in $B_s\to J/\psi f_0$ decay, only the $\bar ss$ and the gluon
component contribute. Moreover, the final state mesons in these
channels are easy to reconstruct and these channels could have
sizable branching fractions. If we use the factorization method,
decay amplitudes are given as
\begin{eqnarray}
 A(\bar B^0\to J/\psi f_0)&=&\frac{G_F}{\sqrt 2} V_{cb}V_{cd}^*
 (2\epsilon^*_{J\psi}\cdot P_B)
 f_{J/\psi} a_2 F_1^{B\to f_0}(m_{J/\psi}^2).
\end{eqnarray}
The Wilson coefficient $a_2$ can be extracted from the $B\to J/\psi
K$ decays~\cite{Amsler:2008zzb}
\begin{eqnarray}
 {\cal B}(\bar B^0\to J/\psi \bar K^0)&=&(8.71\pm0.32)\times
 10^{-4}.
\end{eqnarray}
The branching ratios are roughly predicted as
\begin{eqnarray}
 {\cal B}(\bar B^0\to J/\psi f_0(\bar nn))
 &\simeq&\left\{\begin{array}{c}
      (23 ^{+12}_{-14})
 \times 10^{-6}\;\;\;\;\;{\rm S1}\\
        (10^{+7}_{-5})
 \times 10^{-5}\;\;\;\;\; {\rm S2}
      \end{array}\right.,\\
 {\cal B}(\bar B^0\to J/\psi G)
 &\simeq& (1.5\pm0.5) \times 10^{-6},
\end{eqnarray}
where we have assumed the same $q^2$ dependence for all form factors
and $F_1^{B\to K}(0)=0.3$. The  uncertainties are from the
experimental data for ${\cal B}(B\to J/\psi K)$ and   the $B\to S$
form factors at the $q^2=0$ point.  For the $B_s$ decays, the
branching ratios are comparable with that of $B\to J/\psi K$:
\begin{eqnarray}
 {\cal B}(\bar B_s\to J/\psi f_0(\bar ss))&\simeq & \left\{\begin{array}{c}
     (6.5^{+4.0}_{-4.5})
 \times 10^{-4}\;\;\;\;\;\;\; {\rm Scenario \,1}\\
        (3.5 ^{+2.3}_{-1.4})
 \times 10^{-3}\;\;\;\;\;\;\; {\rm Scenario \,2}
      \end{array}\right.,\\
 {\cal B}(\bar B_s\to J/\psi G)&\simeq& (2.4\pm1.0)\times 10^{-5}.
\end{eqnarray}
Such large branching fractions offer a great opportunity to probe
structures of scalar mesons. With the available data in the future,
the mixing problem between the scalar mesons will be solvable and
the glueball component can be projected out in principle.

\begin{figure}
\begin{center}
\includegraphics[scale=0.4]{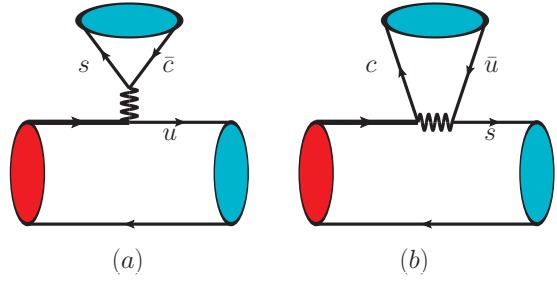}
\end{center}
\caption{Feynman diagrams of $B\to f_0D$ and $B_s\to f_0\bar D$
decays. These channels suffer large uncertainties.}
\label{diagram:FeynBtof0D}
\end{figure}

If the power-suppressed annihilation diagrams are neglected, the
charmful decays of $B$ meson, $B\to f_0D$, can also be used to
constrain the mixing between scalar mesons. For instance in $B^-\to
D_s^- f_0$, the $\bar nn$ and gluon component contribute but the
$\bar ss$ component does not, while in $\bar B_s\to D^0 f_0$, the
$\bar nn$ component will not contribute, as shown in
Fig.~\ref{diagram:FeynBtof0D}. Thus the mixing coefficients can also
be determined if these two channels are experimentally measured. It
is necessary to point out that this method may suffer from sizable
uncertainties of annihilation diagrams~\cite{Wang:2006ria}.

\subsection{Glueball production in $B_c$ decays}

The ordinary light scalar meson is isospin singlet and/or flavor
SU(3) singlet, while the glueball is flavor SU(6) singlet. Therefore
it is difficult to distinguish them by the light $u$, $d$ and $s$
quark coupling. However, the light ordinary scalar meson has
negligible $c\bar c$ component, while the glueball have the same
coupling to $c\bar c$ as that to the  $u \bar u$, $d \bar d$ or $s
\bar s$. A clean way to identify a glueball is then through the
$c\bar c$ coupling to the glueball.

In $B$ decays, the initial heavy meson contains a light quark, thus
contributions of the gluon component always accompany with the quark
content $\bar nn$ or $\bar ss$. It is not easy to isolate the gluon
content. The situation in the doubly-heavy $B_c$ meson is different:
it contains a heavy charm antiquark. The semileptonic $B_c\to
f_0l\bar\nu$ decays would happen only through
Fig.~\ref{diagram:FeynBtof0}(a)(b) and (c) but not through
Fig.~\ref{diagram:FeynBtof0}(d) and (e). The observation of this
decay channel in the experiments will surely establish the existence
of a scalar glueball.   Moreover the CKM matrix element in this
channel is $V_{cb}$, thus the $B_c\to f_0l\bar\nu$ will have a
sizable branching ratio. This channel will depend on the $B_c\to G$
transition form factor which requires the less-constrained $B_c$
meson's light-cone distribution amplitude. But even if we assume the
form factor of $B_c\to G$ is smaller than the $B_c\to \eta_c$ form
factor by one order, branching ratios of $B_c\to G l\bar\nu
(l\bar\nu)$ are suppressed by two orders
\begin{eqnarray}
 {\cal B}(B_c\to Gl\bar\nu)& \sim & 1\% \times 0.01 =   10^{-4},
\end{eqnarray}
where the branching ratio of $B_c\to \eta_c l\bar\nu$  has been
taken as $1\%$. Although   more quantitative results  need   the
precise  $B_c\to G$ form factors, which requires the $B_c$ wave
function with large uncertainties, the order of magnitude of
branching ratio is convincing. This branching ratio is large enough
for the experiments. One only needs to reconstruct the $f_0$ scalar
meson in the final state and also the $B_c$ meson mass in the
intermediate state, so that to make sure that the scalar meson is
produced from two gluons. That experiment is achievable even if the
$f_0$ meson is not a pure glueball, but at least has a large portion
of it. The disadvantage here is the missing neutrino in detector is
hard to be reconstructed in hadronic machines, like LHCb. A future Z
factory is an ideal place for this channel to be measured
\cite{Zfac}.

$B_c\to f_0\pi^-$ is another potential mode to figure out the gluon
content. But in this mode, the $\bar nn$ component also contributes
through the annihilation diagrams. The $b$ and $\bar c$ quark
annihilates and the $d$ and $\bar u$ quark are created. The CKM
matrix element $V_{cb}$ and the Wilson coefficient $a_1$ are the
same with the emission diagram for the $B_c$-to-glueball transition.
The offshellnes of the two internal particles in annihilation
diagrams are of the order $m_{B_c}^2$. The electroweak vertex is the
$V-A$ type and the decay amplitude is proportional to the light
quark mass. Thus the decay amplitudes via annihilation diagram for
the $\bar nn$ component are expected to be suppressed.  As a result,
the $B_c\to f_0\pi^-$ also filters out the gluon component of the
scalar meson as an approximation.

\section{Conclusions}
\label{sec:conclusions}
%

In this work, we have investigated the transition form factors of
$B$ meson decays into a scalar glueball state in the PQCD approach.
Compared with the form factors for the quark content, we find that
form factors for the gluon content have the same power counting in
the expansion of $\Lambda_{\rm QCD}/m_B$ but they are numerically a
little smaller. The pure glueball states can be detected in
semileptonic $B\to Gl\bar\nu$ and $B_s\to Gl^+l^-$ in the future,
since they have a sizeable branching ratio.

If a scalar meson is a mixture of a glueball and an ordinary meson,
we investigate the possibility to extract the mixing mechanism from
semileptonic $B$ decays, such as  $B\to f_0l\bar \nu$ and $B_s\to
f_0l^+l^-$ decays. The nonleptonic $B\to J/\psi f_0$ and $B_s\to
J/\psi f_0$ decays are also analyzed. To be specific, we discussed
the production rates of scalar mesons under two different mixing
mechanisms, and we find that the differences in $B$ and $B_s$ decays
are helpful to distinguish the two mixing mechanisms. To avoid the
interference between the quark and the gluon component, we find that
the $B_c \to f_0 l\bar\nu$ and $B_c\to f_0\pi^-$ will project out
the gluon component of a scalar meson cleanly. Our results can be
generalized to the other glueballs.

\section*{Acknowledgement}

This work is partly supported by National Natural Science Foundation
of China under the Grant No. 10735080, 10625525, and  10847161 and
Natural Science Foundation of Zhejiang Province of China, Grant No.
Y606252. We would like to acknowledge Prof. Hsiang-nan Li for useful
discussions.
\begin{appendix}

\section{hard kernels}


Hard scales and functions in the $B\to G$ form factors are defined
by:
\begin{eqnarray}
 t_a&=& {\rm max}\left[\sqrt {\rho x_1}m_B,1/b_1,1/b_2\right],\\
 t_b&=& {\rm max}\left[\sqrt {\rho x_2}m_B,\sqrt{\rho\bar x_2 x_1}m_B,1/b_1\right],\\
 t_c&=&{\rm max}\left[\sqrt {\rho }m_B,1/b_1,1/b_2\right],\\
 E_a(t_a)h_a&=&\alpha_s(t_a){\exp}[-S_B(t_a)-S_G(t_a)]\left[\theta(b_1-b_2)I_0(\sqrt
 {\rho x_1}m_{B}b_2)K_0(\sqrt
 {\rho x_1} m_{B}b_1)\right.\\
 && \left.+\theta(b_2-b_1)I_0(\sqrt {\rho x_1}m_{B}b_1)K_0(\sqrt
 {\rho x_1}m_{B}b_2)\right] K_0(\sqrt{\rho \bar x_2 x_1}
 m_{B}b_2),\\
 E_b(t_b)h_b&=&\alpha_s(t_b){\exp}[-S_B(t_b)-S_G(t_b)]  \int_0^1 dz \frac{b_1}{2\sqrt
 { Z_1 }} K_1(\sqrt { Z_1 }b_1) ,\\
 E_c(t_c)h_c&=&(E_a(t_a)h_a)|_{x_1\to 1,\bar x_2\to x_2},
\end{eqnarray}
where $Z_1$ is defined as
\begin{eqnarray}
 Z_1=\rho  x_2m_B^2 z+ \rho\bar x_2 x_1 m_B^2(1-z).
\end{eqnarray}


The Sudakov factors are given by:
\begin{eqnarray}
S_B(t)&=&s\left(x_1\frac{m_B}{\sqrt
2},b_1\right)+\frac{5}{3}\int^t_{1/b_1}\frac{d\bar \mu}{\bar
\mu}\gamma_q(\alpha_s(\bar \mu)),\\
S_G(t)&=&s_G\left(x_2\frac{\rho m_B}{\sqrt
2},b_2\right)+s_G\left((1-x_2)\frac{\rho m_B}{\sqrt 2},b_2\right),
\end{eqnarray}
where the explicit form for the  function $s_G(Q,b)$ is
\begin{eqnarray}
s_G(Q,b)=\int _{1/b}^Q\frac{d\mu}{\mu}\left[\ln
\left(\frac{Q}{\mu}\right)A(\alpha(\mu))\right], \;\;\;
A=\frac{\alpha_s}{\pi}C_A,
\end{eqnarray}
with $C_A=3$. Since there is no endpoint singularity in these
diagrams, the contribution from the endpoint region will not be
large and we will neglect the jet function arising from the
threshold resummation. We adopt the one-loop expression of the
running coupling constant $\alpha_s$, when evaluating the above
Sudakov factors

\end{appendix}

\end{document}